\documentclass{emulateapj}

\usepackage{amsmath}
\usepackage{graphicx}
\usepackage{natbib}

\DeclareMathOperator{\divv}{div}
\DeclareMathOperator{\curl}{curl}

\begin{document}

\title{The impact of thermodynamics on gravitational collapse: filament formation and magnetic field amplification}

\author{Thomas Peters\altaffilmark{1,2,3}}
\email{tpeters@physik.uzh.ch}

\author{Dominik R. G. Schleicher\altaffilmark{4}}

\author{Ralf S. Klessen\altaffilmark{1}}

\author{Robi Banerjee\altaffilmark{5}}

\author{Christoph Federrath\altaffilmark{6,1}}

\author{Rowan J. Smith\altaffilmark{1}}

\author{Sharanya Sur\altaffilmark{7}}

\altaffiltext{1}{Universit\"{a}t Heidelberg, Zentrum f\"{u}r Astronomie, Institut f\"{u}r Theoretische Astrophysik,
Albert-Ueberle-Str. 2, D-69120 Heidelberg, Germany}
\altaffiltext{2}{Fellow of the Baden-W\"{u}rttemberg Stiftung}
\altaffiltext{3}{Institut f\"{u}r Theoretische Physik, Universit\"{a}t Z\"{u}rich,
Winterthurerstrasse 190, CH-8057 Z\"{u}rich, Switzerland}

\altaffiltext{4}{Georg-August-Universit\"{a}t, Institut f\"{u}r Astrophysik, Friedrich-Hund-Platz 1, D-37077 G\"{o}ttingen, Germany}

\altaffiltext{5}{Hamburger Sternwarte, Gojenbergsweg 112, D-21029 Hamburg, Germany}

\altaffiltext{6}{Monash Centre for Astrophysics, School of Mathematical Sciences, Monash University, Vic 3800, Australia}

\altaffiltext{7}{Raman Research Institute, C.~V.~Raman Avenue, Sadashivnagar, Bangalore 560080, India}

\begin{abstract}

Stars form by the gravitational collapse of interstellar gas. The thermodynamic response of the gas can be characterized by an effective equation of state. It determines how gas heats up or cools as it gets compressed, and hence plays a key role in regulating the process of stellar birth on virtually all scales, ranging from individual star clusters up to the galaxy as a whole. We present a systematic study of the impact of thermodynamics on gravitational collapse in the context of high-redshift star formation, but argue that our findings are also relevant for present-day star formation in molecular clouds.

We consider a polytropic equation of state, $P=k\rho^\Gamma$, with both sub-isothermal exponents $\Gamma<1$ and super-isothermal exponents $\Gamma>1$. We find significant differences between these two cases. For $\Gamma>1$, pressure gradients slow down the contraction and lead to the formation of a virialized, turbulent core. Weak magnetic fields are strongly tangled and efficiently amplified via the small-scale turbulent dynamo on timescales corresponding to the eddy-turnover time at the viscous scale. For $\Gamma<1$, on the other hand, pressure support is not sufficient for the formation of such a core. Gravitational contraction proceeds much more rapidly and the flow develops very strong shocks, creating a network of intersecting sheets and extended filaments. The resulting magnetic field lines are very coherent and exhibit a considerable degree of order. Nevertheless, even under these conditions we still find exponential growth of the magnetic energy density in the kinematic regime.

\end{abstract}

\maketitle

\section{Introduction}
Star formation occurs under a large variety of conditions, ranging from Milky Way-type molecular clouds, starburst galaxies, environments of supermassive black holes to the primordial Universe. Environmental conditions such as the ambient radiation field, the flux of cosmic rays, and the overall metallicity of the system, determine the thermodynamic response of the gas to compression. At low metallicity, for example, the gas generally cools less efficiently, increasing the characteristic temperatures and changing the slope of the effective equation of state \citep{Omukai05,Omukai12}. In starburst galaxies and close to supermassive black holes, the gas is exposed to UV and/or X-ray radiation fields, providing additional sources of heating, ionization and photodissociation \citep{Maloney96, Kaufman99, Meijerink05}. In starburst galaxies, the high amount of cosmic rays may provide a temperature plateau at high densities, again having a potentially strong impact on the star formation process \citep{Papadopoulos11}. 

Overall, it is evident that the equation of state may change significantly depending on the conditions of the environment. In 3D simulations, the effects of such varying equations of state have however only been explored in rather specific situations. For instance, strong efforts were dedicated to the modeling of primordial star formation \citep{abeletal02, bromlars04, Yoshidaetal08, Clark11, Greif11}. Only a few simulations are available considering star formation in metal-poor environments \citep[e.g.][]{Dopcke12, Glover12, latif12}, and circum-nuclear starburst regions \citep{Klessen07,Hocuk11}. For the modelling of star formation in Milky Way-type molecular clouds, different approaches exist, ranging from piecewise-polytropic equations of state \citep[e.g.][]{Bonnell08, Price08} to the solution of the actual heating-cooling balance \citep{Krumkleinmckee07, petersetal10a, clarketal12}.

In such a variety of different conditions, it is central to obtain a more general understanding of how the equation of state influences stellar birth. For example, \cite{Larson05} provided a detailed discussion of the potential role of the equation of state on the formation of filaments. Based on numerical simulations, \citet{LiKlessenMacLow03} and \citet{Jappsen05}  studied the impact on the overall fragmentation behavior of the gas, but they have not modeled  the process of protostellar collapse and stellar birth on small scales and neglected magnetic fields. Analytic theories suggests a dependence of the local collapse timescale on the equation of state \cite[see, e.g.][]{Yahil83}.

The gas that formed the first stars and protogalaxies initially collapses sub-isothermally \citep{scalo02} and reaches a phase of super-isothermal collapse when the gas becomes optically thick. In between these two density regimes, the behavior can be very complicated and change multiple times, depending on the exact metallicity \citep{omukai10}. In this letter, we systematically explore the impact of different equations of state on protostellar collapse and the star formation process in the context of primordial star formation using 3D simulations. We consider weakly magnetized collapsing minihalos and explore the effect of the equation of state on the general morphology, the turbulent properties, and the structure and growth of the magnetic field. Particular attention is paid to magnetic field amplification processes, which may play an important role even at early cosmic times \citep{Schleicher10c, Sur10, Federrath11, Schober12}.  

In the following, we outline our numerical approach in Section~\ref{method}, and present our results in Section~\ref{results}. We summarize our findings and discuss their implications in Section~\ref{discussion}.

\section{Numerical Simulations}\label{method}

We carry out numerical simulations with the adaptive mesh (magneto-)hydrodynamics code FLASH \citep{fryxell00}. We use a positive-definite, MUSCL-Hancock, Riemann scheme HLL3R \citep{waagan11} to solve the magnetohydrodynamic equations. Because our aim is to study the details of the dynamical evolution during protostellar collapse, we adaptively increase the numerical resolution as the density increases and ensure that the local Jeans length is always resolved with at least 64 cells. Previous studies showed that at least 32 grid cells per Jeans length are required to resolve vorticity and turbulent magnetic-field amplification on the Jeans scale \citep{Sur10,Federrath11}, so we will safely capture these effects in our simulations. However, the amplification of small-scale magnetic fields and vorticity occurs on the viscous scale, where the non-linear term in the induction and vorticity equations lead to an exponential amplification \citep[e.g.,][]{federrath11b}. The viscous scale in real systems, however, is typically much smaller than can be resolved with any current simulation technique. Thus, even though the total kinetic energy on the Jeans scale is converged with a Jeans resolution of 32 cells \citep{Federrath11}, vorticity production on smaller scales is not. The amount of vorticity produced in simulations like the ones discussed below increases with the numerical resolution, as the effective Reynolds number increases at the same time \citep{Sur10,Federrath11,Turk12}. Furthermore, baroclinicity (misaligned pressure and density gradients), which is absent in our barotropic simulations, can additionally contribute to vorticity production. Thus, the magnetic field and vorticity produced in our simulations is a lower limit of the actual amount produced in real systems.

For our initial condition we use the central region of a primordial minihalo representative of those in which the first stars form. The gas in such a minihalo consists purely of Hydrogen and Helium, and at the central densities considered here ($10^{-15}\,$g\,cm$^{-3}$) follows an almost isothermal equation of state ($\Gamma\sim 1.1$) in the purely metal-free case. To allow a direct comparison between the simulations and a systematic study, we adopt the same initial conditions for all equations of state. The gas distribution is smooth and centrally concentrated, and the gas temperature is high, meaning that the internal turbulence is sub- to transsonic. Intrinsic magnetic fields in primordial minihalos are thought to be weak, but we show here that very small initial magnetic fields are exponentially amplified on timescales shorter than the collapse.

Our initial condition was extracted from the cosmological simulations of \citet{Greif11}. We select a typical minihalo and extract its central region on scales of the Jeans length, corresponding to $8000\,$AU, with a total mass of $101.3\,$M$_\odot$ and a mean temperature of $710\,$K. The initial peak density of the simulation is $2.8 \times 10^{-15}\,$g cm$^{-3}$. The original simulation was carried out using the Lagrangian mesh code Arepo \cite[][]{Springel10} and remapped onto a cartesian grid with a resolution of $256^3$ grid cells to generate the initial conditions for the resimulations discussed here. 

The initial, turbulent magnetic field has a power-law spectrum, $P_B(k)\propto k^{3/2}$ on large scales \citep{Kazantsev68}, typical of the small-scale dynamo \citep{Brandenburg05} and peaks on a scale corresponding to 20 grid cells. The peak scale was set such, because \citet{Federrath11} found a range of scales corresponding to 20--30 grid cells, on which the magnetic field grows fastest due to small-scale dynamo action. Since we want to address potential formation of coherent large-scale magnetic fields in this letter, we chose the lower limit in order to ensure that we start with a truly small-scale field. Below this resolution limit, the spectra drop with $P_B(k)\propto k^{-4}$. The simulation results are, however, insensitive to the details of this drop in the dissipation range. This initial, turbulent magnetic field serves as a sensible initial condition for a field that was previously amplified by small-scale dynamo action in a collapsing halo \citep{Federrath11,Xu11,Turk12}. In all simulations we have set the amplitude of the random magnetic field to an initial value of $10^{-9}\,$G, which is too small to reach the regime where the small-scale dynamo saturates during the collapse but allows us to study its exponential growth in the kinematic regime of the turbulent dynamo.

To cover the range of possible thermodynamic response behavior of star-forming gas at different metallicities and in different environments, we adopt a simple polytropic equation of state, $P=k\rho^\Gamma$. We have run  four collapse calculations with different polytropic indices $\Gamma = 0.7, 0.9, 1.1, 1.2$. A value  $\Gamma < 1$ corresponds to a strong cooling regime as provided, e.g., by molecular or atomic line emission at all metallicities or the coupling to dust at low metallicites \cite[e.g.][]{Klessen12}, while $\Gamma > 1$ corresponds to a heating regime as realized when the gas becomes optically thick or couples to the dust at high metallicities \cite[e.g.][]{Glover12a}. 

Our primary motivation is the study of gravitational collapse and stellar birth in the early Universe, where magnetic fields are initially weak and their subsequent amplification is of interest. However, the insights we gain from altering the equation of state apply to a range of different scenarios
and may also be relevant to the dynamics of nearby star-forming clouds in the Milky Way at present days.

\section{Results}\label{results}

\begin{figure*}[t]
\centerline{\includegraphics[width=360pt]{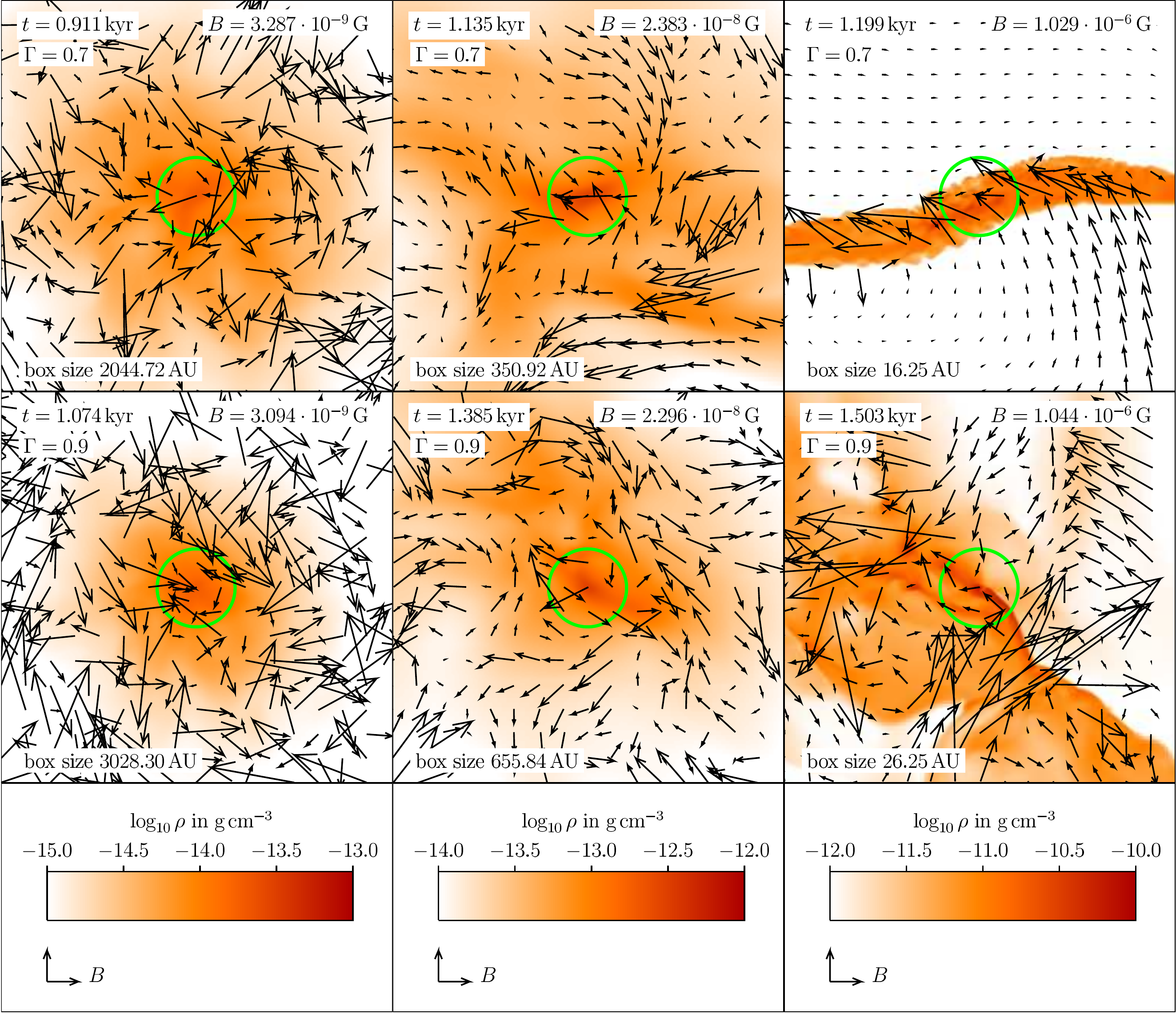}}
\vspace{5pt}
\caption{Magnetic field and density structure for the four runs as function of time. Rows are different polytropic indices ($\Gamma = 0.7, 0.9$, from top to bottom), columns are different times (time advances from left to right). The magnetic field vectors have been rescaled for plotting by $\rho^{2/3}$. The field strength mentioned in the images corresponds to an arrow of the length given in the legend. The green circle marks the Jeans volume.}
\label{fig:magneticfielda}
\vspace{5pt}
\centerline{\includegraphics[width=360pt]{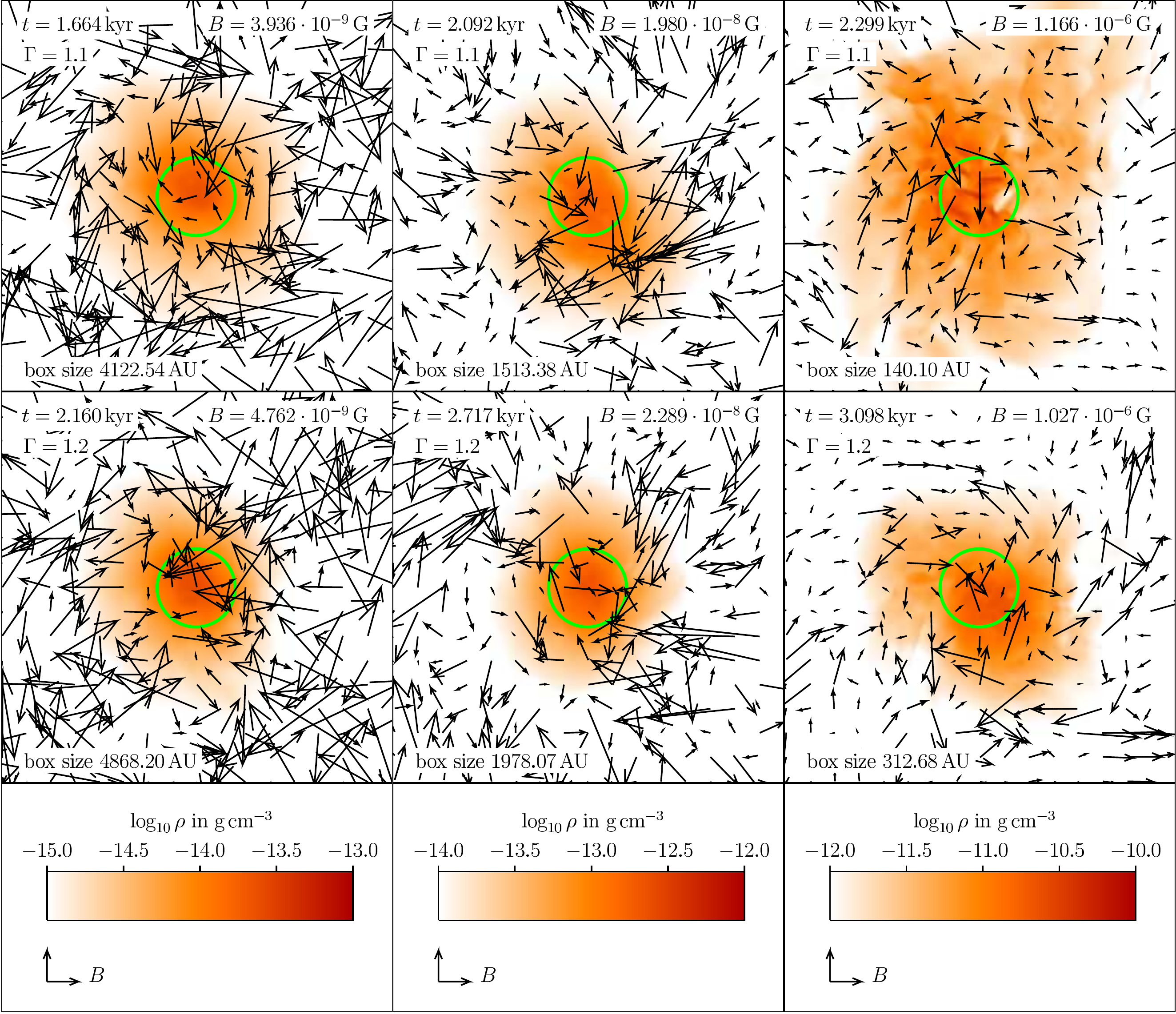}}
\vspace{5pt}
\caption{Magnetic field and density structure for the four runs as function of time. Rows are different polytropic indices ($\Gamma = 1.1, 1.2$, from top to bottom), columns are different times (time advances from left to right). The magnetic field vectors have been rescaled for plotting by $\rho^{2/3}$. The field strength mentioned in the images corresponds to an arrow of the length given in the legend. The green circle marks the Jeans volume.}
\label{fig:magneticfieldb}
\end{figure*}

The collapse dynamics is remarkably different depending on the polytropic exponent. To illustrate this point, we show in Figures~\ref{fig:magneticfielda} and \ref{fig:magneticfieldb} density slices and magnetic field vectors for the four simulations (rows) at three different times (columns). The density structure and magnetic field topology changes significantly when the polytropic exponent shifts from super-isothermal ($\Gamma > 1$) to sub-isothermal ($\Gamma < 1$).

In the super-isothermal case, strong pressure gradients are able to build up. They slow down the gravitational collapse and the flow tends to virialize in the central core. The accretion onto this core is able to generated strong turbulence \cite[e.g.][]{Klessen10,Federrath11}. In the kinetic weak-field regime studied here, the magnetic field is strongly tangled and efficiently amplified via the small-scale dynamo on timescales corresponding to the eddy-turnover time at the viscous scale \citep{Kazantsev68, Schober12}. In contrast to that, sub-isothermal collapse exhibits a markedly different dynamical behavior. In the absence of strong (stabilizing) pressure gradients, gravitational contraction proceeds much more rapidly. As a consequence, the flow develops very strong shocks, which create a network of intersecting sheets and extended filaments rather than one central virialized core. The resulting magnetic field lines are very coherent and considerably more ordered than in the case $\Gamma > 1$.

\begin{figure*}[t]
\centerline{\includegraphics[width=360pt]{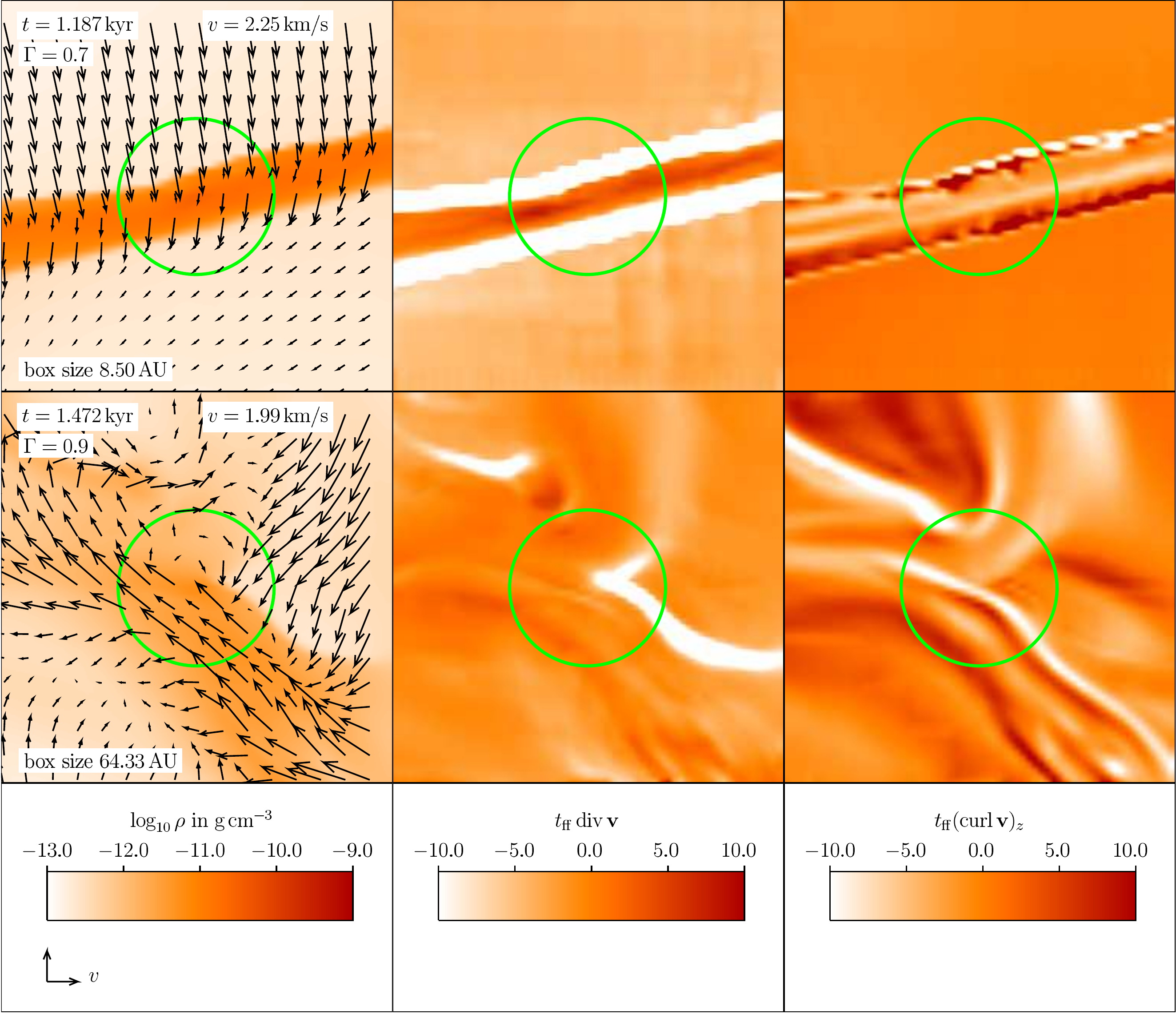}}
\vspace{5pt}
\caption{Velocity structure for the runs with $\Gamma = 0.7, 0.9$ (from top to bottom) at the time shown in the first column. The columns show density, $t_\mathrm{ff} \divv \mathbf{v}$ and $t_\mathrm{ff} (\curl \mathbf{v})_z$, respectively. The arrows are velocity vectors. The velocity mentioned in the images corresponds to an arrow of the length given in the legend. The green circle marks the Jeans volume.}
\label{fig:velocityfielda}
\vspace{5pt}
\centerline{\includegraphics[width=360pt]{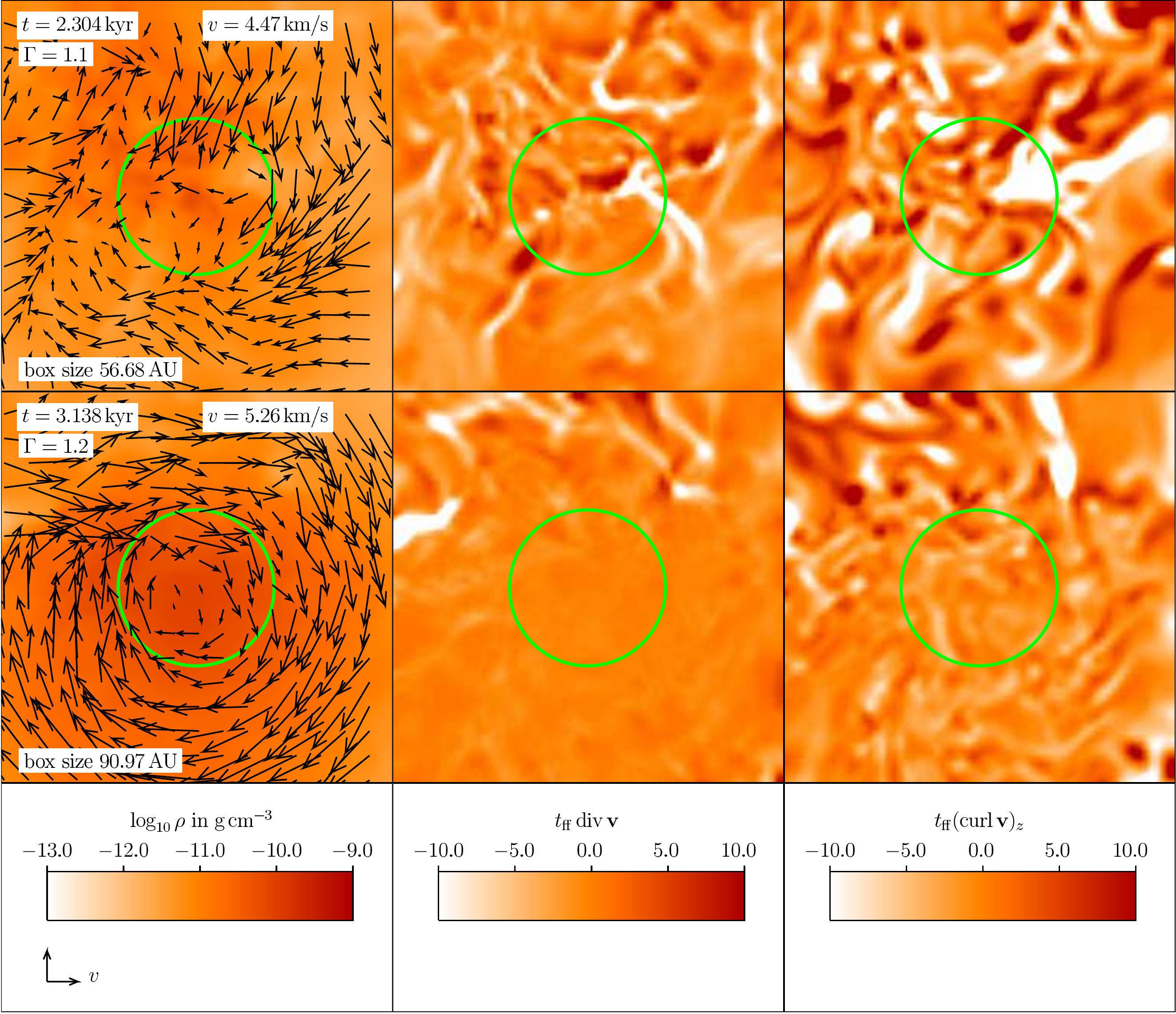}}
\vspace{5pt}
\caption{Velocity structure for the runs with $\Gamma = 1.1, 1.2$ (from top to bottom) at the time shown in the first column. The columns show density, $t_\mathrm{ff} \divv \mathbf{v}$ and $t_\mathrm{ff} (\curl \mathbf{v})_z$, respectively. The arrows are velocity vectors. The velocity mentioned in the images corresponds to an arrow of the length given in the legend. The green circle marks the Jeans volume.}
\label{fig:velocityfieldb} 
\end{figure*}

The two different regimes can also be seen in the velocity structure. Figures~\ref{fig:velocityfielda} and \ref{fig:velocityfieldb} show the divergence of the velocity field $\divv \mathbf{v}$ and the $z$-component of the vorticity $(\curl \mathbf{v})_z$, both normalized by the local freefall time $t_\mathrm{ff}$, for one snapshot of each simulation. The simulations are compared at similar mean densities in the Jeans volume. Clearly, the velocities have significantly more small-scale structure in the turbulent core of the super-isothermal than in the sub-isothermal cases. It is  interesting to note that even for $\Gamma < 1$ we find both compressive and solenoidal turbulent modes close to the curved shock structures (see Fig.~\ref{fig:velocityfielda}), but very little velocity structure in the infalling regions. This behavior can be explained by vorticity production in curved shocks \citep{mee06,federrath11b} and by viscous interactions at density gradients, efficiently operating even in regimes of primarily compressive driving of turbulence, as during gravitational contraction studied here.

\begin{figure}[t]
\centerline{\includegraphics[width=180pt]{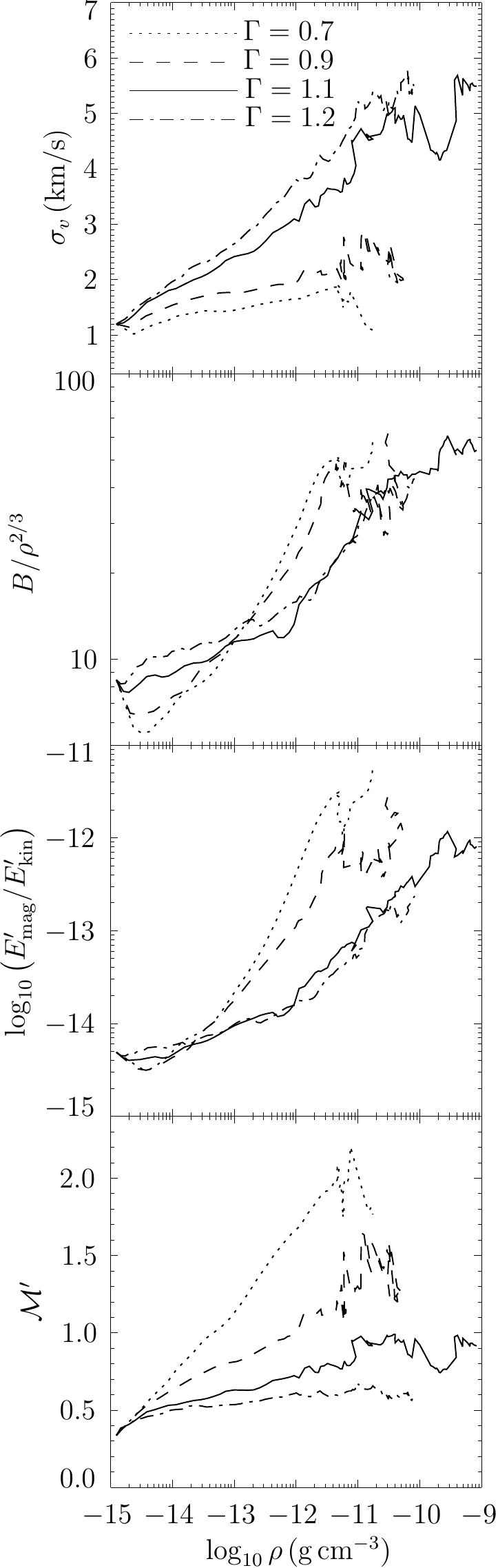}}
\caption{Evolution of the turbulent velocity dispersion, effective magnetic field amplification, ratio between the small-scale magnetic field and the turbulent kinetic energy, and the turbulent Mach number within the Jeans volume for the collapse simulations with $\Gamma = 0.7, 0.9, 1.1, 1.2$.}
\label{fig:panels}
\end{figure}

We now turn our attention to the magnetic field amplification. Indeed, the effect of the two different dynamical regimes is clearly visible in Figure~\ref{fig:panels}. We plot  the turbulent velocity dispersion $\sigma_v$, the magnetic field amplification normalized by the contribution from flux-freezing $B / \rho^{2/3}$, the ratio of the fluctuating parts of the kinetic energy and magnetic energy $E_\mathrm{mag}' / E_\mathrm{kin}'$, as well as the turbulent Mach number $\mathcal{M}'$ for the four simulations. All quantities are averaged over the Jeans volume during the collapse and plotted against the mean density in the Jeans volume. Contributions from a mean velocity or magnetic field have been subtracted to calculate the fluctuating components. The growth of the turbulent velocity dispersion is much weaker for the sub-isothermal collapse than in the super-isothermal case. Regarding the evolution of the magnetic field, the ratio $B/\rho^{2/3}$ is increasing for all four simulations, confirming the presence of additional dynamo action beyond pure compression. We note that we find stronger magnetic field amplification for the sub- than for the super-isothermal situation. However, the interpretation of this result is difficult, because for $\Gamma < 1$ the collapse is filamentary and so the normalization by a factor $\rho^{2/3}$ does not compensate completely for the contribution from flux-freezing. 

\section{Discussion and conclusions}\label{discussion}

We performed a set of four high-resolution numerical simulations, where we studied gravitational collapse and magnetic field amplification for four different thermodynamic regimes. We find that the dynamics of gravitational collapse and consequently the star formation process depends significantly on the thermal evolution of the gas \cite[e.g.][]{Larson85}. If the gas is able to heat up during compression, i.e.\ for an effective $\Gamma>1$, we find the formation of a central virialized core, leading to the production of turbulent motions and highly tangled magnetic field structures \cite[see also][]{Sur10, Federrath11}. Under sub-isothermal conditions ($\Gamma < 1$), the formation of such cores is suppressed. The collapse proceeds much faster and the flow develops strong shocks. These create a network of intersecting sheets and extended filaments. In this case, vortices still form in the vicinity of the curved shocks and at the intersection of multiple shocks \citep{sun03}, but are absent in the infalling region. 

We find that magnetic fields are amplified in the super-isothermal as well as in the sub-isothermal region. The local magnetic field strength is increased both by compression, which is more efficient within the shocks, and by the presence of turbulent motions. The turbulent magnetic field amplification can be described in terms of the Kazantsev theory, and has been shown to depend on the square root of the Reynolds number for Kolmogorov turbulence and on the cubic root of the Reynolds number for Burgers turbulence \citep{Kazantsev68, Subramanian98, Schober12b}. The transsonic simulations presented here lie in between these two extreme cases. Because of the large Reynolds number of astrophysical gases, the inferred amplification timescales are typical of order of, and under most conditions significantly shorter than, the local free-fall time \citep{Federrath11,Schober12}. We therefore expect the presence of magnetic fields close to the saturation level for a wide range of $\Gamma$ (even for $\Gamma<1$) quite independent of environmental conditions. In particular, we argue that magnetic fields must not be neglected when studying stellar birth in the high-redshift Universe. 

We expect the equation of state to play a major role also under different conditions, for instance in the observed filamentary structure of nearby star-forming clouds \citep{Andre10,Arzoumanian11}. We note that the interstellar medium (ISM) in the solar neighborhood at number densities $n$ in the range between $10$ and $\sim10^4\,$cm$^{-3}$ lies in a strong cooling regime with  $\Gamma \approx 0.7$ \citep{Larson85}, where cooling is provided by the line emission mostly of C$^+$ and CO \citep{Glover12a}, before coupling to dust results in $\Gamma \approx 1.1$ for densities above $n \sim 10^5\,$cm$^{-3}$. It is well established that the supersonic turbulence ubiquitiously observed in the ISM by itself will lead to a highly filamentary morphology \citep[e.g.][]{maclow04}. Here we want to draw attention to the fact that also the thermodynamic response of the gas plays an important role in shaping the physical characteristics of the observed filaments \cite[see also][]{Larson05}. We argue that the cooling processes associated with the formation of molecular clouds are a key to understand the observed morphological structure of nearby star-forming regions.

\section*{Acknowledgements}

We thank Thomas Greif for providing the initial conditions for our simulations and Simon Glover for discussions on non-isothermal effects in gas clouds. We also thank the anonymous referee for useful comments that helped to improve the paper.

T.P. acknowledges financial support as a Fellow of the Baden-W\"{u}rttemberg Stiftung funded by their program International Collaboration II via contract research grant P-LS-SPII/18 and through SNF grant 200020\textunderscore 137896. D.R.G.S. thanks the Deutsche Forschungsgemeinschaft (DFG) for funding through the SPP 1573 {\em Physics of the Interstellar Medium} (project number SCHL~1964/1-1) and the SFB 963 {\em Astrophysical Flow Instabilities and Turbulence}. R.S.K. and R.J.S acknowledge support from the DFG via the SPP 1573 (grants KL~1358/14-1 \& SM~321/1-1) and via the SFB 881 {\em The Milky Way System} (sub-projects B1, B2, B3). R.B. acknowledges funding from the DFG via the grant BA 3706/1-1. C.F. acknowledges funding provided by the Australian Research Council under the Discovery Projects scheme (grant DP110102191). S.S. thanks RRI for support during the course of this work.

We acknowledge computing time at the Swiss National Supercomputing Centre, the Leibniz-Rechenzentrum in Garching (Germany), the NSF-supported
Texas Advanced Computing Center (USA), and at J\"ulich Supercomputing Centre (Germany). The FLASH code was in part developed by the DOE-supported Alliances
Center for Astrophysical Thermonuclear Flashes (ASCI) at the University of Chicago.

\end{document}